# Game Design for Blockchain Learning

Diogo Cortiz
Network Information Center (NIC.br)
Pontifical Catholic University of São Paulo (PUC-SP)
São Paulo, Brazil
dcortiz@pucsp.br

Newton Calegari
W3C Brazil Office
Network Information Center (NIC.br)
São Paulo, Brazil
newton@nic.br

Fabiana Oliveira
CS:Games Study Department
Pontifical Catholic University of São Paulo (PUC-SP)
São Paulo, Brazil
moliveira.fabiana@gmail.com

Daniel Couto Gatti
Computer Science Department
Pontifical Catholic University of São Paulo (PUC-SP)
São Paulo, Brazil
daniel@pucsp.br

*Abstract:* **Blockchain is a new technological approach that has gained popularity on the market due to its application in several areas such as education, health, security, and smart cities, among others. However, understanding how blockchain works is not easy at first, especially for non-technical people, because it relies on a non-trivial computational process. We have developed a game board - called Blocktrain - whose game mechanics are based on the blockchain processing model. This game gives people the opportunity to learn key blockchain concepts while playing. In this paper, we describe the game design process and assessment of the game as pedagogical instrument.**

## I. INTRODUCTION

Blockchain is a new technological approach that is gaining popularity in the market due to its possible applications in several areas such as education, health, security, and smart cities, among others. It is a kind of distributed system that has its origin in the financial sector, specifically in the technology that supported the creation of the Bitcoin cryptocurrency.

For decades, researchers in the field of computer science and cryptography have worked on research projects to develop a digital currency, but they have always faced the challenge of the double-spending problem, which happens when someone can spend the same amount of money twice. The nature of digital technologies leads to this scenario, because it is easy to reproduce.

However, researches in these areas have taken a new direction after Satoshi Nakamoto (actually a hacker's pseudonym) published an article in which he describes a data-processing approach to a virtual currency that would not only solve the double-spending problem, but also bring security and transparency for transactions [1]. Nakamoto addressed the problem by proposing an electronic system that uses a network composed of different nodes in which all transactions are verified and recorded using cryptographic algorithms. In this case, each node within the network checks and validates transactions with no central authority [2, 3].

The system works as follows: a transaction (money transfer, in the case of Bitcoin) is stored in a temporary block, which remains open for a specific amount of time to receive transactions. Once the time limit has been reached, the block is closed, and is not able to receive new transactions for then on, being sent to the network to be processed. This is called "mining", a validation phase in which all the nodes of the network, through an encryption system, can perform some algorithms to find the right valid key. Once the block is validated, it is propagated by the network so that all nodes have the same information. Then the next block will reference the previous one, and so on, in a dynamic in which a block always refers to the previous one, thereby creating a chain.

It is important to mention that Nakamoto never used the word 'blockchain' in his original paper [1], but researchers and programmers understood that this unprecedented computational process could be applied in several areas in which transparency and reliability are important. Based on the way the technology works, they coined the name 'blockchain' to describe this approach of data processing that could be incorporated in different platforms and technologies.

Nowadays, there are many expectations that blockchain can revolutionize various segments, such as the economy, education, and even our political system [4]. The idea is that blockchain can be used to store different types of data that require security, transparency and immutability (such as health data, legal data, educational data, in addition to only money transfer data).

Many organizations and governments are already looking into how to develop and adopt blockchain to provide better services. It is important that decision-makers and people from different areas know about blockchain so that they can envisage ways in which technology can be applied in their contexts.

However, understanding blockchain is not easy at first, especially for non-technical people, because it is a non-trivial computational process. Considering this, we developed a board game called Blocktrain, whose game mechanics was based on the blockchain processing model. The aim of this game is to help people understand key blockchain concepts in a game-based approach.

The motivation to develop a game to facilitate the understanding of blockchain arose from a need to teach about this technological approach during the School of Internet Governance, organized by the Internet Steering Committee in Brazil (cgi.br). The School of Internet Governance follows the guidelines of EuroSSIG (European Summer School on Internet Governance) and has an audience of people from different walks of life, including judges, prosecutors, lawyers, teachers, researchers, executives and decision makers. As it has a broad audience with little technical knowledge about blockchain, we suggested a new approach to explain the basic concepts of blockchain through a game. The game was entirely developed by an interdisciplinary team with different backgrounds (computer scientists, game designers, artists and storytellers) from different labs, research centers and universities.

The game was successfully used during the School of Internet Governance in Brazil. After that, it was taken up as a pedagogical instrument for blockchain learning in several Brazilian universities, in both in undergraduate and graduate programs.

In this paper, we present the entire game design of Blocktrain and discuss its development process – how we have incorporated the main concepts of blockchain in a board game and the main challenges we have faced during this process. We also discuss the role of game-based learning and present a survey – methodology and results – carried out to evaluate the game as a pedagogical instrument.

## II. BLOCKTRAIN: GAME DESIGN

The concept of game design is still deeply discussed, especially with regard to its scope. In this work, we rely on the definition that game design is the process of choosing what a game should be like [6]. This is a broad concept, but is important because it does not restrict the process to only one area, such as aesthetics, rules, stories or even programming.

In this section, we describe the narrative developed for the game, the aesthetic design used in the cards, and all the rules. It is important to mention that we decided to develop a board game to help us create a visual metaphor for the blockchain process, where each board represents a block full of transactions.

The boards should be placed in the centre, with the players all around them. The players represent the nodes in a network, and can always keep an eye on the boards. The game design aimed to consider the following premises:

- all players should play the role of a network node, which could make transactions and validate the blocks (mining);
- the game should have a style that is more collaborative, rather than competitive;
- the game should address some of the main features of blockchain: distributed processing, validation/mining and immutability of data.

We have started developing the game design by devising a narrative to contextualize and create an atmosphere for the players. The narrative was created to explicitly show of the problems that blockchain can address: manipulation of information and divergent data.

The gameplay begins with players reading the following narrative, presented in a newspaper format:

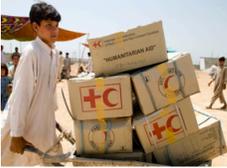

Fig. 1. Newspaper introducing game narrative

In this narrative we present a situation that can happen in real life and for which blockchain can be a solution: the existence of two documents that seem authentic, but with divergent information. This narrative stimulates the debate about how the blockchain can help in the immutability of data and the creation of a chain of trust, creating a game atmosphere for players.

Inspired by this narrative, we have designed a simple goal for the game: players must collaboratively fill train wagons with three kinds of supplies (drinking water, food, and medicine) to be sent to the refugee camp. After filling each wagon, the contents of the wagon must be validated, and each player will be able to do this in his or her round. The train only departs when at least five wagons are filled and duly validated. The collaborative appeal is to have all players participating in the game by filling in and validating the wagons.

The game has three main components:

- Boards – representation of the train wagons which must be filled with Supplies Cards by the players

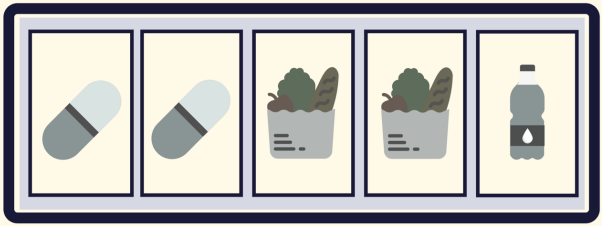

Fig. 2. Example of a board (train wagon)

- Supplies Cards (drinking water, food and medicine) – cards that players will use to fill the wagons

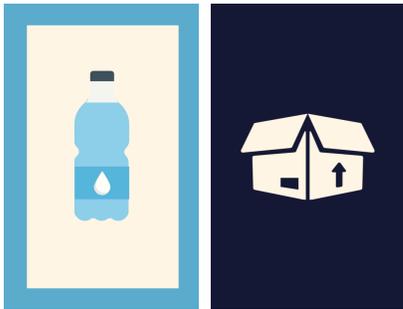

Fig. 3. Example of a drinking water supply card (front and back)

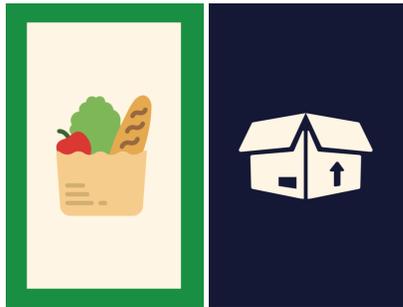

Fig. 4. Example of a food supply card (front and back)

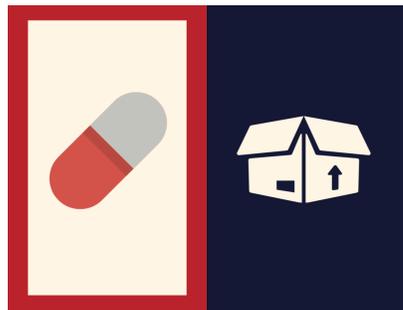

Fig. 5. Example of a medicine supply card (front and back)

- Validation Cards - cards with a combination of supplies (drinking water, food and medicine) that will be used by players to validate a wagon

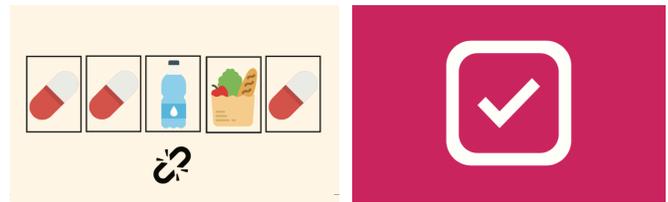

Fig. 6. Example of a validation card (front and back)

Rules are also elements that are part of the mechanics of the game. They define how actions happen and what players can and cannot do. Below we present the rules that are part of this game design.

- The maximum number of players is five.

- Each player receives four (4) Supplies Cards and 1 (one) Validation Card. The remainder of the cards are grouped into two different Draw Piles, one with Supplies Cards and the other with Validation Cards.

- For each round, a new board (train wagon) is placed on the table, and the player who dealt the cards begins filling in the Wagon. The game shall rotate clockwise, and the fill should always be from left to right.

- The player must fill the supply space in the board (train wagon) with exactly the same supply card (drinking water, food or medicine). Each player can play once per round. If the player has the card with the right supply, he/she can fill in the board. If he/she does not, he/she can pick up one (1) card from the Draw Pile (Supplies Cards). The game then moves to the next player.

- When all positions on the board (Wagon) are filled, the validation phase starts. The last player who played a Supply Card starts the validation round. If the player has the right Validation Card (which must have the exact sequence of the supplies present on the board), he/she can validate the wagon by positioning it on the board. Each player may pick up one card from the Draw Pile (Validation Cards) by round. The game moves to the next player until someone can validate the wagon.

- Once the wagon is validated, a new board is laid on the table. Supply Cards and Validation Cards are shuffled and redealt to players. Each player should receive four (4) Supplies Cards and one (1) Validation Card.

- The game ends when five (5) wagons (boards) are duly filled and validated. Based on the narrative, the train with the supplies then sets off for the refugee camp. It is a collaborative game, so there is no winner. Players collaborate to achieve the goal: fill five (5) wagons (boards).

It is known that games as metaphors can be effective when used as learning instruments [6]. With this game mechanics, we tried to elaborate metaphors for some blockchain concepts, in order to facilitate their understanding.

One approach was the development of mechanics in which each player plays the role of an active node of the network, being able to transact (in this case fill the wagon with Supply Cards), and also participate in the validation phase, as if they were the miners (in this case, finding the correct Validation Card).

The wagon (board) assumes the role of a block of a blockchain, which can store several transactions and, necessarily, needs to be validated. By connecting one wagon with the other, we create a visual metaphor of a train to explain the basic architecture of a blockchain platform, which helps in its understanding, moving from a purely conceptual explanation to an illustration.

In the following image we can see how the wagons (boards) are arranged after the end of the game, creating a visual metaphor for the blockchain architecture.

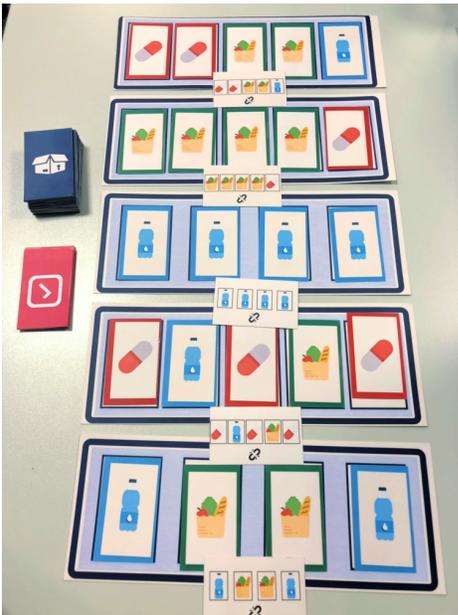

Fig. 7. Example of boards' layout after the end of the game

The main purpose of this game is to facilitate the understanding of some features and functionalities of the blockchain, which are not easily learned from an initial explanation. The distributed architecture of the technology is the basis of its operation and we seek to portray it in the game through a dynamic in which each player acts as a node of the network, being able to insert transactions and validate the blocks.

Thus, it is possible for players to notice that data is maintained by all of them - that data in blockchain does not belong to just one node. Transparency and trust are made evident as all players (nodes) have the same knowledge about what is happening with the supplies (data) present in the train (blockchain).

The immutability of information is perhaps the main advantage in a blockchain platform, a mechanism that ensures that no information could be edited or deleted. This functionality was represented in the game in its own dynamics: it is only possible to add data (in this case, insert Supply Cards) in the current block and before the validation. Once the block is validated, it can no longer be changed.
If a player tries to make a change, the others players will easily notice and the order of supplies on the board will be different from the Validation Card, invalidating the block.

The mining process is one of the key mechanisms of the blockchain because it ensures the security of transactions. This is represented in the game during the validation phase. In this step, each player plays the role of a miner in search of the key that validates the block.

The computational effort required to validate a *hash* in blockchain technology is transposed into the players' dedication to always picking up cards from the Draw Pile (Validation Card) in order to find the correct key to validate the wagon board (Block). Every round, they should pick up a new Validation Card from the pile until someone finds the correct one. This part of gameplay is based on randomness and an element of luck.

The game design described in this section is the first version of the game. We are investigating how it can be improved based on the evaluation as presented in the next section. The game has already been used as a pedagogical instrument in several universities in Brazil, with satisfactory results. It is also important to mention that the game design is freely available based on creative commons licences, by which any person or institution can use and improve it.

### III. EVALUATION: BLOCKTRAIN AS A GAME-BASED LEARNING APPROACH

Game-based learning is an approach that seeks to use certain game principles and apply them to get users involved in the process of knowledge acquisition. The dynamics and the play element of games stimulate students to get even more involved in the learning process. Game-based learning is not just about creating games for students for the purpose of fun: it is designing learning activities that can present concepts guiding them toward a final goal [8].

The game was designed to be used as an object to support blockchain learning, making it easier to understand the less trivial concepts of blockchain. The intention is not to make the player learn all blockchain concepts directly while playing, but that the metaphors developed in the game may help the player in the assimilation of knowledge when there is an explanation about blockchain afterwards.

Metaphors are important for making connections between the abstract and the concrete, and games are known to be an excellent tool for creating metaphors in computing learning. With this type of content, situations and abstract problems are represented in concrete, visual and interactive situations [9, 10].

In our case, the Blocktrain game was used as a learning instrument, for starting a blockchain class. As the game has an independent and self-contained narrative, students are invited to play it without any previous explanation about blockchain. After the game, the specialised teacher/instructor can make references to the game during the theoretical explanation of the blockchain in order to facilitate understanding through metaphors.

This game was used during The School of Internet Governance in Brazil and in some undergraduate and graduate courses in Brazilian universities. We have conducted a study to identify the perception and the satisfaction of players regarding Blocktrain as a fun game as well as a learning tool.

*A. Method*

The main objective of the study was to evaluate the students' perception about how the game helps to understand the basics concepts of blockchain. Our proposal was to evaluate the game based on two main dimensions: the game design (fun, rules and time duration) and the game as a pedagogical instrument (how the game helps the learning of blockchain).

We have prepared a survey with five (5) questions based on a five-point Likert type response scale (totally agree – agree – neutral– disagree – totally disagree), and one (1) open-ended question.

To analyze the five (5) questions based on the Likert Scale, we used data analysis and descriptive statistics. For the open-ended question, we used open coding and axial coding [11]. During the open coding, we have identified concepts and discovered their properties in the data. During the axial coding, we have created connections between the codes, grouping them based on their properties to generate categories.

We conducted this research on two occasions: the first during the School of Internet Governance, with a universe of 30 people; and the second in a graduate class of an important business school in São Paulo, with an attendance of some 20 people.

On both occasions, we started the activity with students playing the game followed by a theoretical explanation about blockchain. Only at the end of the class were the students invited to respond to the questionnaire. In total, 32 people of different ages and backgrounds took part in the survey and answered the questionnaire.

*B. Results*

The first dimension analyzed in this research is game design. In this step, we focus on three main components, according to the perception of the players:

1) fun: whether players think Blocktrain is a fun game or not. Our intention is to collect players' perceptions of the game in general terms.

2) duration: in this item we would like to know if the duration of the game is ideal (not too short, yet not too exhaustive), especially for a class context

3) easy learning: in this item we seek to understand if the rules of the game are easy to understand.

In the first item, we asked the players if they thought the game was fun:

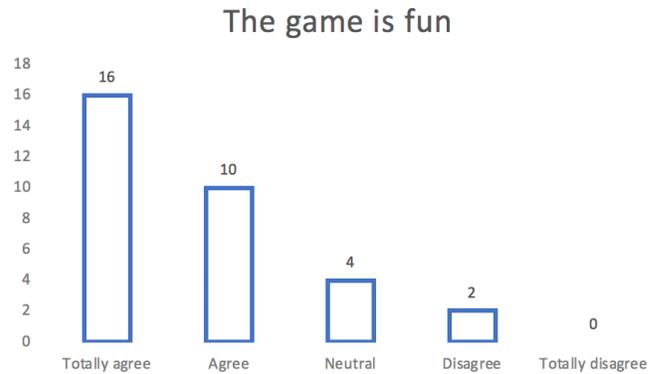

Fig. 8. Distribution of responses to the statement:  the game is fun

As Fig. 8. shows, it turned out that most players perceived Blocktrain as a fun game in general terms. Twenty-six players (81.25%) agreed that the game is fun, and 16 (50%) agreed totally. Only 2 of the players (6.25%) disagreed. This is a positive evaluation for a game with the main function of a teaching tool, rather than just being a fun activity.

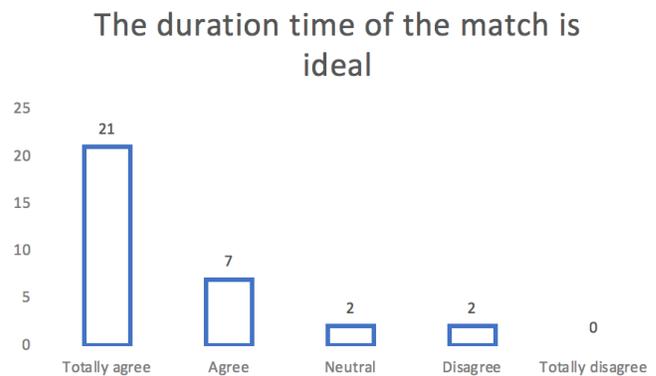

Fig. 9. Distribution of responses to the statement: the duration of the game is ideal

The typical duration of the game is approximately 20 minutes. This is the time it usually takes for players to achieve the goal of the game: to fill and validate at least 5 (five) wagons.

As we can see in Figure 9, most respondents believe that this is a suitable time for the game. Twenty-eight (87.5%) players agreed this time is ideal, and 21 (65.6%) totally agreeing with the statement.

Based on our empirical experience after we had played the game several times, our opinion is also that this is an adequate duration for the game, especially if it is used as a warm-up activity.

In our opinion, a game focused on learning should be easy to understand the first time. If it takes a lot of time to understand the rules, this can cause confusion and players may lose interest in the game. In this project, our goal was to balance a mechanics of blockchain metaphor with easy rules. However, this was a challenge during the development process, and we had to redesign the game a few times to get to the final game design version presented in this paper.

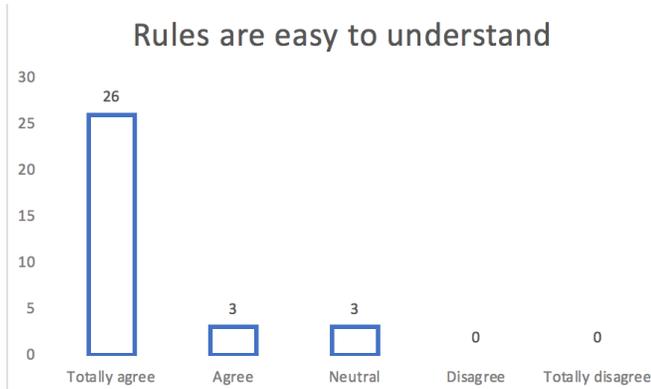

Fig. 10. Distribution of opinion responses to the statement: the rules are easy to understand

According to the respondents in the survey, as shown on Figure 10, we can conclude that we have been able to develop a game with simple and easy rules. Indeed, 29 of the players (90.7%) agreed that the rules are easy to understand, with 26 (81.3%) fully agreeing with this statement. It is important to note that no player disagreed with the statement.

It is important to mention that, based on our experience after playing this game with different classes, we have never noticed any evidence suggesting that the game is difficult to understand. At all times, the experience was fluid, easy, and fun for the players.

These three survey items based on a Likert scale are part of the game design evaluation. The main mission of the project was to develop a game with focus on the pedagogical approach, but we have always looked for bringing fun to game design, so Blocktrain could exist as a game *per se* and not just as a boring and discouraging pedagogical instrument. The results are satisfactory in this sense: the game is fun, easy to understand and with an ideal duration.

Despite the satisfactory evaluation of the players, we can make some comments that we noticed after the game had been played several times. The game does indeed fulfil its role as a fun and entertaining object; however, we do realise that it becomes less exciting after being played more often.

In the very design of the game, we can observe that its mechanics is based on randomness and an element of luck, which limits the decision making of the player. When played for the first time, the game is a novelty and there are many aspects constituting an element of surprise which peters out as the game is played more often. Our future work aims to develop some new game dynamics in which the player has to build strategies and make more decisions.

The second dimension analysed in this research is the function of the game as a pedagogical instrument. We have dedicated two (2) items in the survey and one (1) open-ended question to evaluate these aspects. The purpose of this evaluation is to identify whether the game helped in understanding blockchain, and how this happened.

The first item in the survey in this dimension is to assess whether players have been able to identify blockchain elements in the game. Thus, we can assess whether the metaphors used in the game have effects or not.

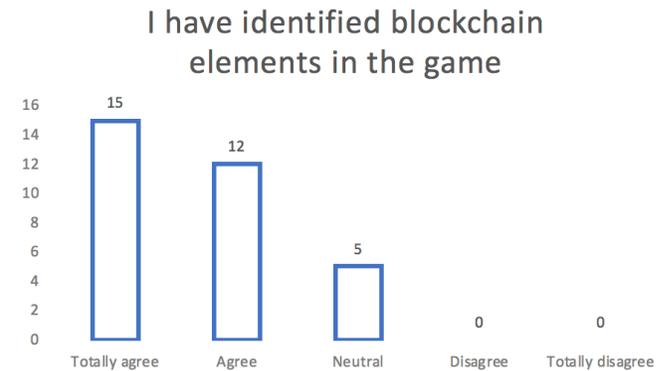

Fig. 11. Distribution of opinion responses to the statement: I have identified blockchain elements in the game

According to Figure 11, the vast majority of respondents said they were able to identify blockchain elements in the game. Twenty-seven players (84.4% of the total) agreed that there are elements of blockchain in the game, with 15 (46.9%), fully agreeing with the phrase.

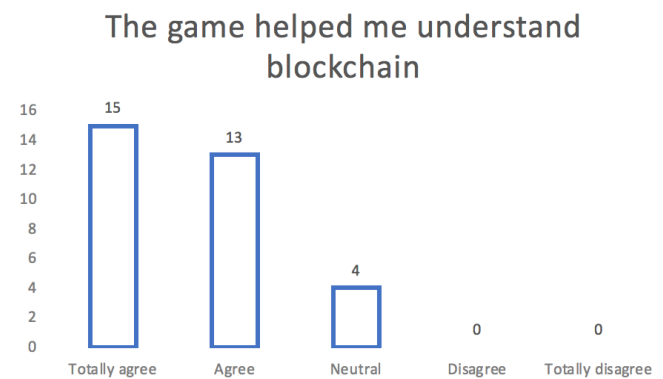

Fig. 12. Distribution of opinion responses to the statement: the game helped me understand blockchain.

The second item of this dimension aims to assess whether the game helped to understand the blockchain in general terms. The results are very positive. Based on Figure 12, twenty-eight players (87.5%) said the game helped them to understand blockchain, and 15 (46.9%) players totally agreed with the statement. There was no disagreement with the statement

In order to further investigate which blockchain elements are the most identified in the game by the players, we have created an open-ended question "What are the main elements of Blockchain found in the game?". To analyse the responses, we used open coding and axial coding [11] as explained in the method section, which resulted in three main classifications presented in Figure 13.

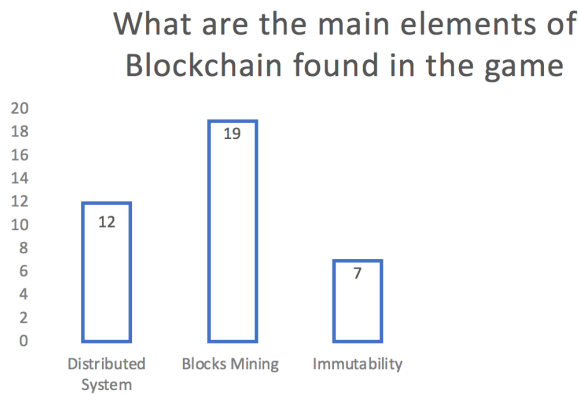

Fig. 13. Answers to the question: What are the main elements of Blockchain found in the game?

During data analysis, we have noticed that the players could recognize three key elements of blockchain: blockchain as a distributed system, blocks mining, and immutability. We have created those three classifications based on players' responses.

For example, respondent C said that "*I could figure out a distributed network among all the players when I play*". For this statement, we have coded it as a mention to "*distributed network*" code.

Respondent F stated that blockchain "*increases security in transactions, where the validation control of blocks is not in the hands of only one node*". We have coded this statement as mention to "*decentralization*" once the respondent said the process is not in the hands of only one node, but we also coded this statement as a mention to "block validation".

Respondent Y stated that he/she found "*the existence of several nodes in the network and the process of blocks validation and data immutability*". For this statement, we have coded it as a mention to "*distributed network*", "*block validation*" and "*data immutability*".

Based on those three examples of codification, we have grouped "*distributed network*" and "*decentralization*" codes to "*Distributed System*" classification, while we have grouped "block validation" to "*Blocks Mining*" classification and "*data immutability*" code to "*Immutability*" classification.

After running this coding process throughout our whole sample, we have identified 12 occurrences in the responses that point to the distributed system of the blockchain as an element present in the game.

This characteristic is evident in the game, as each player plays the role of a node of the network, being able to transact and validate the blocks.

We have identified 19 occurrences for mining process. It is the most important feature of blockchain and we have made an effort during the game design phase to make it clear in the gameplay. We have implemented the validation phase where each player can act as a miner of a blockchain platform. This data suggests the result was achieved.

Finally, we have identified seven occurrences in the responses for the immutability element found in the game. This is a feature that is inherent in blockchain architecture. We planned to deal with this as an emerging phenomenon in gameplay in a non-obvious way. By the dynamics of the game, players should recognize that they can only add things, but never change or delete something.

Based on this analysis, we can see that the game has the potential to achieve our initial objectives. Some characteristics of blockchain are more easily found in the game, such as block mining, but even immutability is a phenomenon that could be explored during the learning process.

## IV. DISCUSSION AND CONCLUSION

Blockchain is a technological approach that has emerged from the mechanism used by Bitcoin to validate all transactions, which ensures security and avoids double spending. With the popularity of Bitcoin, it was soon perceived that this computational approach could be used to solve several technological problems that demand security, reliability and often the immutability of data.

Thus, the use of Blockchain began to be studied in a plethora of different segments, such as financial, industrial, health, and the education sector, among others. This technological approach is not easy to understand at first, especially by non-technical people, because it is a technology that adopts a non-trivial architecture of distributed processing and encryption. However, it is important that people in all areas understand it, so they can propose to use cases in their own contexts.

The Blocktrain game was designed as a board game, to facilitate its use in classrooms. The strategy adopted was that of creating metaphors of blockchain elements within the game. Another reason for choosing a board game model rather than a digital game is that a board game model could create a visual metaphor for the blockchain.

In this paper, we describe the game design in detail. It is important to mention that the first step we adopted during the project was the development of a narrative that would show a problem that the blockchain could solve. In this case, we developed a narrative in which there is a discrepancy of data in a transport document for donations sent on to a refugee camp. The narrative is important to create a context for the class, to motivate the players and encourage a game atmosphere.

After the narrative, we then started to design the mechanics and rules of the game, based on the story. At this stage, we faced the challenge of balancing the fun with the didactic appeal of the game. Sometimes we made changes to the mechanics of the game to make it more fun, but these changes caused the loss of the blockchain metaphor in the game. We prototyped three versions of the game design up to the final version presented in this paper.

In this research we also prepared a survey to evaluate the project in two dimensions: the game design and the purpose of the game as a pedagogical instrument. During the analysis of the results, it was possible to identify that the game has a game design quite balanced. The game is fun, it's easy to understand (simple rules), and the duration of a game is ideal.

The research also evaluated the game as a pedagogical instrument. From the beginning of the project, the objective was to facilitate the understanding of blockchain concepts through metaphors in a game. The results of the survey carried out with the players reveal that the game has helped in learning, mainly in identifying the main blockchain concepts: distributed systems, mining/validation and immutability.

The preliminary results of this research show that the main objectives of the project have been achieved. A game with a consistent game design was developed for a pedagogical purpose: to help learning blockchain concepts. This paper is about a work in progress, but in future work, we intend to modify the game design to add more strategy elements to the game so the players can make more decisions in the gameplay experience, thereby reducing the randomness and the element of luck present in the current version. We also aim to apply this survey to a larger number of participants to increase the survey sample.